# A Nanocar and Rotor in One Molecule

*Kwan Ho Au-Yeung[1], Suchetana Sarkar[1], Tim Kühne[1], Oumaima Aiboudi[2], Dmitry A. Ryndyk[3, 4], Roberto Robles[5], Nicolas Lorente[5, 6], Franziska Lissel[2], Christian Joachim[7], Francesca Moresco[1]\**

[1]Center for Advancing Electronics Dresden, TU Dresden, 01062 Dresden, Germany

[2]Leibniz-Institut für Polymerforschung Dresden e.V., 01069 Dresden, Germany, and Faculty of Chemistry and Food Chemistry, TU Dresden, 01062 Dresden, Germany

[3]Institute for Materials Science, TU Dresden, 01062 Dresden, Germany

[4]Theoretical Chemistry, TU Dresden, 01062 Dresden, Germany

[5]Centro de Física de Materiales CFM/MPC (CSIC-UPV/EHU), 20018 Donostia-San Sebastián, Spain

[6]Donostia international physics center, 20018 Donostia-San Sebastián, Spain

[7]GNS & MANA Satellite, CEMES, CNRS, 29 rue J. Marvig, 31055 Toulouse, France

ABSTRACT. Depending on its adsorption conformation on the Au(111) surface, a zwitterionic single-molecule machine works in two different ways under bias voltage pulses. It is a unidirectional rotor while anchored on the surface. It is a fast-drivable molecule-vehicle (nanocar) while physisorbed. By tuning the surface coverage, the conformation of the molecule can be selected to be either rotor or nanocar. The inelastic tunneling excitation producing the movement is investigated in the same experimental conditions for both the unidirectional rotation of the rotor and the directed movement of the nanocar.

KEYWORDS. Molecule-rotor, Nanocar, inelastic tunneling electrons, molecular manipulation, scanning tunneling microscopy, density functional theory

Understanding the working principles of synthetic molecular machines is of fundamental importance to develop nanoscale mechanical molecular devices able to, for example, perform calculations,[1] store energy,[2] or produce work.[3] While the first molecular machines have been explored as ensembles in solution[4-6], it is now crucial to control the mechanics of always one and the same molecule.

Specific examples of controlled rotations or translations of a molecule on a surface under the tip of a scanning tunneling microscope (STM) have been recently reported[7] and it is now known that in a STM junction the tunneling electrons can release energy to specific conformational or vibrational molecular degrees of freedom, being in some cases able to trigger rotations[8-12] or translations[13-15]. However, a general understanding of the physical mechanisms inducing controlled and directional movements on a surface is still lacking, which would allow the rational design of single-molecule machines towards specific applications.

In order to enable the rotation of a single molecule, a stable rotational axle and the restriction of the lateral diffusion on the supporting surface are the prerequisites. Different strategies of constructing molecular axle via, *e.g.*, positioning the molecule-gear on a defect,[16] triggering on-surface reactions,[17-18] and building molecule-stators[19-20] have been reported. On the other hand, controlling the translation on a surface is a parallel goal of research. Mobile physisorbed molecules with legs, wings or wheels have been proposed to investigate translations induced by the STM tip.[13-14, 21-26] Furthermore, the first two editions of the international Nanocar Race have provided examples of controlled lateral motions on a surface, induced by inelastic tunneling and electric field.[27-28]

In the STM tunnel junction, the initial excitation mechanism for rotation and translation is a priori the same. However, a rotation requires a strong local molecule-substrate interaction providing a rotational axle, while a nanocar moves fast if its adsorption is weak. It is therefore unusual that the same molecule presents both functions. Here, we present such a molecule. Thanks to its internal charge separation and to the possibility of chemisorb or physisorb on the Au(111) surface, the zwitterionic molecule studied in this work can rotate (without translating) in one direction or can translate (without rotating) depending on its conformation on the surface. Such dual functionality allows the comparison of inelastic tunnelling electrons excitation for a controlled translation or a unidirectional rotation on the same molecule and in the same experimental conditions.

RESULTS AND DISCUSSION

**The molecule**

2-(2-Methoxyphenyl)-1,3-dimethyl-1*H*-benzoimidazol-3-ium iodide (o-OMe-DMBI-I)[29] and further DMBI derivatives have been first synthetized to be employed as air-stable n-type dopants in organic electronic devices.[30-31] It was however soon observed that decomposition reactions occur during the doping process, with the halide separation or the splitting of a methyl group.[32] When sublimating o-OMe-DMBI-I in ultra-high vacuum (UHV), the molecule undergoes a dissociation reaction where the methoxy group is cleaved and methyl iodide is released, yielding the zwitterionic phenolate DMBI-P.[11] In a recent study, we showed that on Au(111) DMBI-P works as a single-molecule rotor reliably converting STM voltage pulses into controllable unidirectional rotation.[11] Here, we investigated the zwitterionic phenolate to make the methoxy cleavage superfluous, and elongated the rotating part to a naphthyl, obtaining 2-(1,3-dimethyl-1*H*-naphtho[2,3-d]imidazol-3-ium-2-yl)phenolate (DMNI-P) presented in Figure 1a. The chemical structure of o-OMe-DMBI-I, DMBI-P and DMNI-P are compared in Figure S1.

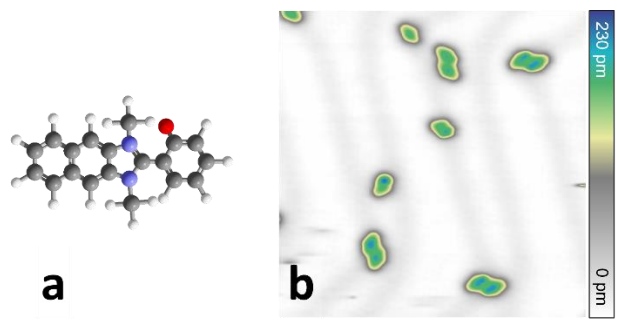

**Figure 1.** (a) Chemical structure of the zwitterionic DMNI-P (color code: C = gray, H = white, N = purple, O = red); (b) STM image of single DMNI-P molecules and dimers after sublimation on the Au(111) surface. Image size: 40 nm x 40 nm, I = 5 pA, V = 0.2 V.

We thermally deposited the DMNI-P molecules under UHV conditions ($T_{evap}$ = 225 °C) on the clean Au(111) surface kept at room temperature. STM experiments were performed after cooling the sample to T = 5 K. The molecular coverage on Au(111) was selected by varying the sublimation time.

A representative overview STM image (Figure 1b) shows a moderate coverage of about 0.1 ML. Single molecules and dimers can be observed where the majority occupies the kinks of the Au(111) reconstruction sites. At lower coverage only single molecules are present on the surface, while by increasing the coverage the relative number of dimers increases, becoming dominant at 0.2 ML (see Figure S2).

**Single-molecule rotors**

We firstly consider the low coverage case and the isolated single adsorbed molecules (Figure 2). A typical STM image of an isolated DMNI-P molecule (Figure 2a) shows the naphtyl unit at the tail of the molecule as well as the widening of the structure roughly in the middle, as the two methyl groups of the imidazole unit appear wider in the STM image because they are not planar.

The molecule is anchored via the phenolate (purple cross).

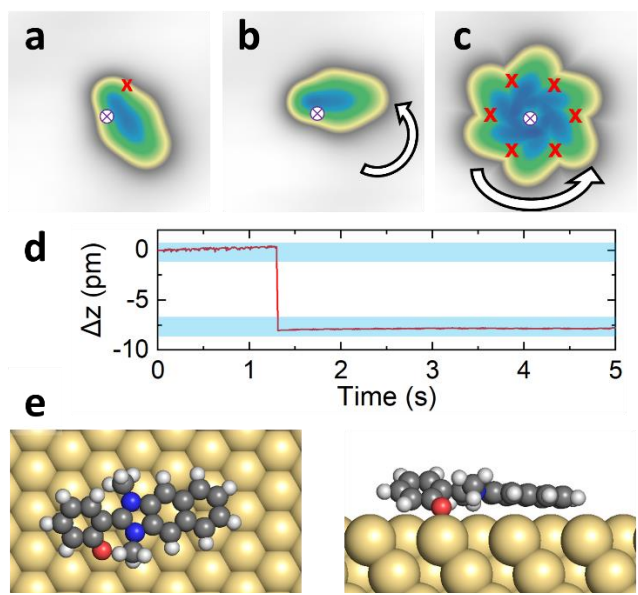

**Figure 2.** Anchoring and rotation of an isolated DMNI-P. (a) STM image of the molecule anchored at the position of the purple cross; a voltage pulse is applied at the position of the red cross. (b) STM image of the same DMNI-P rotor after the voltage pulse, showing a one-step rotation ($\Theta = 60°$; CCW); the anchoring point (center of rotation) is indicated in purple. (c) By combining the images of the six stations rotating in one direction (CCW), the common anchoring point is identified (indicated by a purple cross). (d) Example of a tip-height time plot $\Delta z(t)$ measured during a pulse. A voltage pulse was applied under $I = 250$ pA and $V = 0.5$ V for 5 sec in constant current mode. STM images (3 nm x 3 nm) were taken at $I = 5$ pA and $V = 0.2$ V. (e) Adsorption geometry of the anchored DMNI-P in top and side view respectively, calculated by DFT.

If we position the STM tip apex on a specific area of the molecule (red mark of Figure 2a) and apply a voltage pulse, we are able to controllably induce a one-step rotation around the anchoring point. The tip apex height is chosen to avoid a direct mechanical interaction with the molecule (4 – 5 Å above the surface). A typical voltage pulse is applied at $I = 250$ pA and $V = 0.5$ V for 10 seconds in constant current mode. After the voltage pulse, we found that the molecule of Figure 2a rotated 60° counterclockwise (CCW) to the next stable station (Figure 2b). By applying a succession of pulses following the position of the molecule, we observe the complete CCW rotation of the molecule over six stations of 60° each, corresponding to the stable adsorption positions on Au(111). The center of rotation (axle) can be experimentally determined by superimposing the STM images of six stations from a complete rotation cycle (Figure 2c). During each voltage pulse, a sudden jump of the tip-height signal is recorded (Figure 2d), indicating the time lapse for a one-step rotation. The adsorption geometry calculated by density functional theory (DFT) is shown in Figure 2e confirming that the single DMNI-P chemisorbs on Au(111) by charge back-donation at the oxygen position. Simulated STM images are reported in Figure S4.

After having performed 251 pulses on different rotors, we observed that the molecule rotates in steps of 60° and in the same direction in 99% of the cases. On a given molecule rotor, the entry port for the electron tunnelling allowing a one-step rotation is not located at the anchoring point (Figure 2a), making necessary a re-position of the tip after each pulse (Figure 2c). Our molecule-rotor is chiral after adsorption on the surface. We observe that DMNI-P molecules with opposite chirality rotate in opposite direction under the same voltage pulse: Isomers as the one shown in Figure 2, which we call left-DMNI-P, always rotate CCW by positive voltage pulses. Negative bias pulses drive the rotor in the opposite direction.

Lateral manipulation at low STM junction tunnelling resistance (*e.g.* 10 mV, 1 nA) has been used to confirm that the rotors are anchored on the surface (Figure S5). A STM pushing[33] does not break the Au-O bond and no lateral movement of the molecule can be observed. More interestingly, we are able to flip a molecule-rotor changing its surface chirality, therefore inverting its rotational direction without breaking the Au-O bond. An example of such flipping is reported in Figure 3.

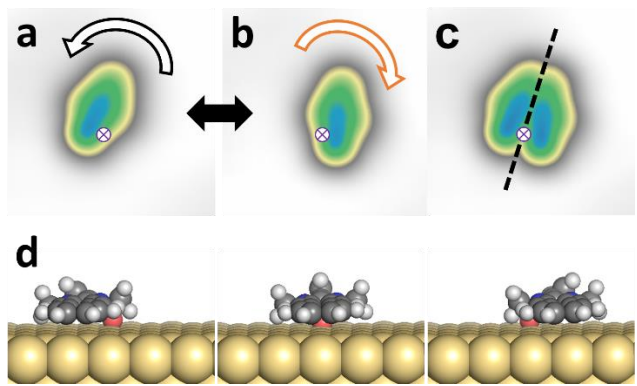

**Figure 3.** Flipping of the chiral rotor molecule leads to the change of rotation direction (CCW → CW). (a) Initially, by applying positive voltage pulses (V = 0.4 – 1.0 V, I = 500 – 700 pA), the rotor rotates in the CCW direction. Then, a stronger pulse is applied above the molecule (V = 3.5 V, tip-surface distance = 10 Å; position not shown). (b) After the pulse, subsequent manipulations show that the rotation direction has been changed from CCW to CW. (c) Combining the two images (a) and (b) by img c = max(img a, img b) gives rise to mirror images, indicating that the molecule is symmetrically flipped to the opposite side without losing the initial anchoring point (indicated by purple marks). (d) Sequence of three DFT calculated adsorption geometries illustrating the flipping.

Figure 3a shows the STM image of a left-DMNI-P, which rotates CCW by positive voltage pulses (typically 0.4 – 0.7 V). After applying a stronger voltage pulse (1.5 – 3.5 V) or by lateral manipulation, the molecule moves (flips) to the opposite side of the bond without changing its anchoring point (Figure 3b). The STM image after flipping clearly shows that the molecule presents now the right-hand chirality. By positive voltage pulses of typically 0.4 – 0.7 V, the unidirectional rotation in the opposite direction is observed, *i.e.* the flipped molecule rotates now clockwise (CW). The flipping is graphically represented in Figure 3c after superimposing the two images of Figure 3a and b. As one can see, the two mirror images are visible, sharing the same anchoring point. Notably, the flipping is reversible and can be repeated several times. DFT calculations show that the flipping of the chirality is due to the rotation of the phenolate-head of the molecule (sequence of Figure 3d), while the naphtimidazole remains adsorbed on the surface shifting laterally. An animation of the simulated conformational change is shown in the Supporting Information.

**Dimers and their separation**

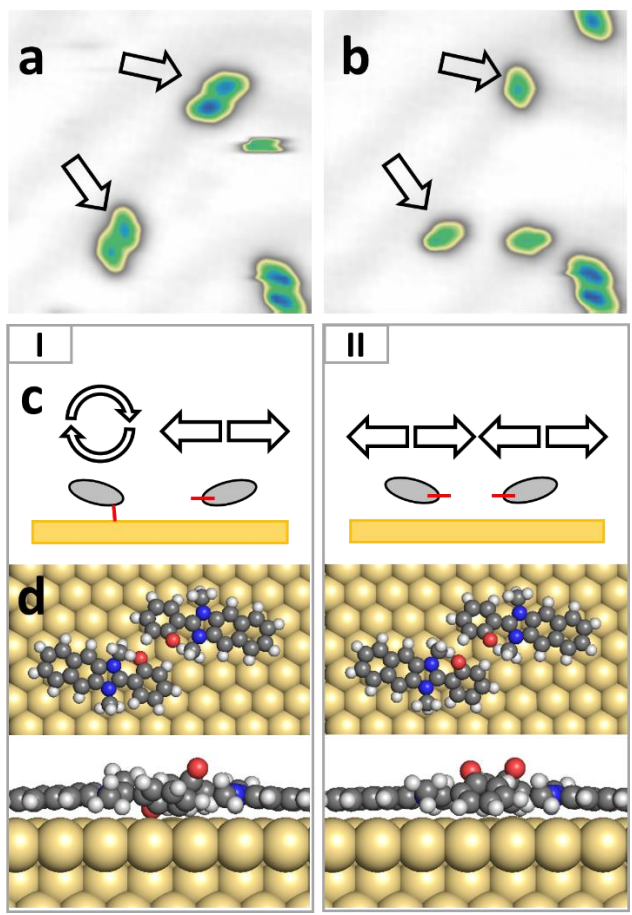

**Figure 4.** STM images (10 nm x 10 nm) (a) before and (b) after applying a voltage pulse (5 V, outside of the image). Dimers can be isolated into single ones (arrow). (c) Schematic diagrams of cases **I** and **II** after separation of dimers. Case **I** shows a molecule rotating and the other translating, and **II** shows both translation behaviors. The red lines illustrate the oxygen moieties undergoing chemisorption or physisorption on the surface. STM images were taken under (a) I = 20 pA, V = 0.2 V, (b – c) I = 10 pA, V = 0.2 V. (d) Adsorption geometry calculated by DFT for **I** and **II**, respectively.

By increasing the molecular coverage on the surface (see Methods section for details), we observe an increasing quantity of molecular dimers. Figure 4a presents the STM image of a few dimers, formed by identical DMNI-P molecules. We can separate the dimers into single DMNI-P molecules by voltage pulses. Typically, we applied a voltage pulse outside of the scan area (*e.g.* 5 V, 160 pA; laterally about 8 nm away from the target). An equivalent separation can be obtained by applying weaker pulses (*e.g.* ~2 V, 5 nA) directly above the dimer (Figure S6). After separation (Figure 4b), we observe in the STM image single adsorbed molecules, where at least one of both molecules has experienced a lateral displacement and, instead of rotating around a fixed point, can be laterally moved on the surface by voltage pulses, behaving like a nanocar. Considering the n = 41 performed dimer separations, about 40% of them result into two separated nanocars and 60% in a nanocar and a rotor.

The schematic Figure 4c summarizes the two observed adsorptions after dimer separation: Case **I** shows that the dimer forms a rotor and a nanocar by voltage pulses; case **II** represents that both the molecules can translate on the surface after separation, *i.e.* they are nanocars. The red lines illustrate the phenolate undergoing chemisorption (Au-O bond) or physisorption on the surface. Figures 4d show the corresponding DFT calculated adsorption geometries for dimer **I** (rotor – nanocar) and dimer **II** (nanocar – nanocar). As one can see, the dimers are formed via hydrogen bonds between two zwitterions, where the O atoms point to the second molecule, hence contributing to the dimer stability. Our calculations show that dimers can be formed already in the gas phase with a bonding energy of 0.2 eV.

The separation of the dimers allows us to contemporary observe the voltage-pulse induced rotation and translation of DMNI-P molecules under exactly the same experimental conditions like surface preparation and tip (see also the movie in the Supporting Information).

**Nanocars**

In the following, we consider the driving of the nanocars on the Au(111) terraces. Figure 5a shows the STM images of a single DMNI-P nanocar after dimer separation.

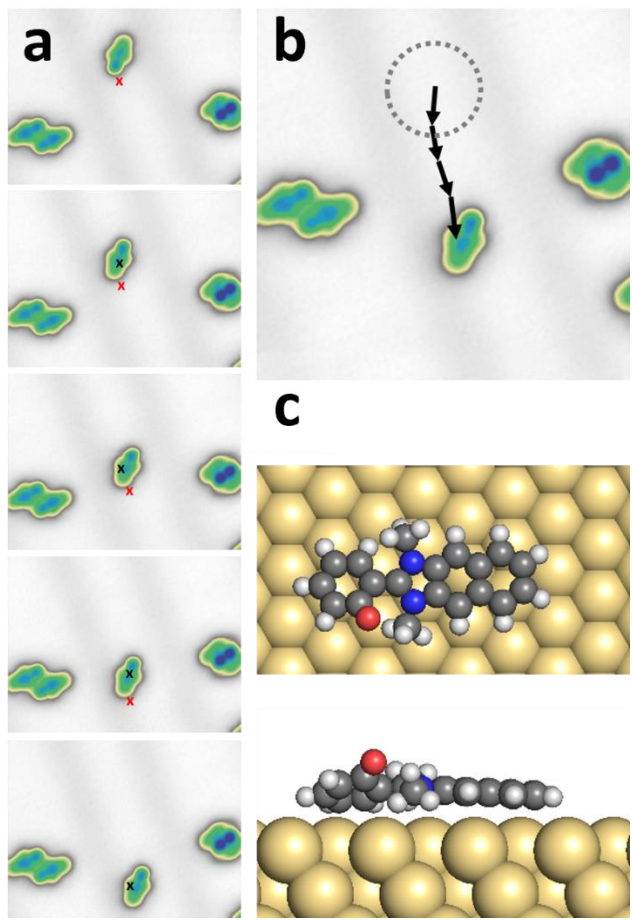

**Figure 5.** Manipulation of the DMNI-P nanocar. (a) Controlled lateral translation starting from the top to the bottom of the 10 nm x 10 nm area can be observed, where the molecule moves towards the pulsing positions. The red marks indicate the position of the tip during the voltage pulses before the displacement, respectively (V = 0.5 – 0.8 V). (b) Trajectory of the molecular movement from (a). STM images (10 nm x 10 nm) were taken under I = 21 pA and V = 0.25 V. (c) Adsorption geometry of the DMNI-P nanocar in top and side view respectively, calculated by DFT.

Its appearance is very similar to the rotor discussed above. A manipulation sequence is shown in Figure 5a and 5b. We position the STM tip apex at the periphery of the molecule (see Figure S8) and apply a voltage pulse. DFT calculations of the adsorbed geometry propose a second stable conformation (Figure 5c) where the O atom is rotated slightly away from the surface and the molecule is physisorbed, interacting with the Au(111) surface by van der Waals interaction. The reaction path (NEB DFT) calculations show that the energy of the car configuration is a little larger than the energy of the rotor configuration but the energy barrier between these configurations is about 0.4 eV making impossible switching from one to another at low temperatures.

We studied the movement of the DMNI-P nanocars applying voltage pulses with the STM tip, similar to the case of the rotors. Figure 5a shows a sequence of a nanocar laterally translating on the Au(111) surface by positioning the tip near to the nanocar (red marks). By applying positive voltage pulses, the molecule is attracted by the tip. Similar to the case of the rotors (Figure 2d), a sudden jump of the tunneling current signal indicates that a hopping event occurs. In this particular sequence, the nanocar has travelled about 4 nm with four individual pulses (Figure 5b). We performed 379 pulses on different nanocars, observing that at positive bias the molecule translates towards the tip position with very high precision, independently on its chirality. On the other hand, the molecule moves in random directions and away from the tip position at negative bias (Figure S11). Typically, the nanocars can be driven at $V \geq 0.4$ V and few hundreds pA. At higher bias voltages ($V > 1.5$ V), the distance between tip and nanocar for an effective translation event can be increased to about 2 nm, resulting in a displacement of 2 nm (attractive to the tip). Please consider that in these conditions resonant tunneling into electronic excited states (see Figure S9) and electrostatic effects are starting to dominate.

**Comparison car-rotor**

Thanks to the possibility of having both chemisorbed and physisorbed conformations of DMNI-P on Au(111), we have compared the unidirectional rotation and translation in the same experimental conditions, applying the same voltage pulses. We induced 251 rotation events and 379 translations on the same surface with the same STM tip and analyzed the data statistically (see Methods for details). We extracted the reaction time of each event from the z(t) plots (see Figure 2d), calculated the average rate R, and plotted the experimental yield, or action spectra,[34] (Y = R/(I/e): motion probability per tunneling electron) for rotors and cars, respectively (Figure 6).

Figure 6 shows the electronic yield for DMNI-P rotations (top plot) and translations (bottom plot) depending on the bias voltage for different currents. In both cases, the different adsorption position of the molecule on the Au(111) surface reconstruction causes a spread of the experimental data, which is obviously stronger in the case of the mobile nanocars.

At voltages higher than 400 mV, the yield for rotation and translation are very similar, and a nearly constant yield of about $5 \times 10^{-10}$ events/electron is observed in both cases. For rotations, a threshold voltage at about 400 mV is visible in the upper plot, which can be explained by the onset of the C-H stretch mode of the methyl group at the opposite side of the Au-O bond, in agreement with the case of the shorter DMBI-P.[11] Fitting the experimental action spectra by the theory of Ref.[34] (continuous lines in the upper panel of Figure 6), we obtain a threshold at 370 meV (see Methods for details).

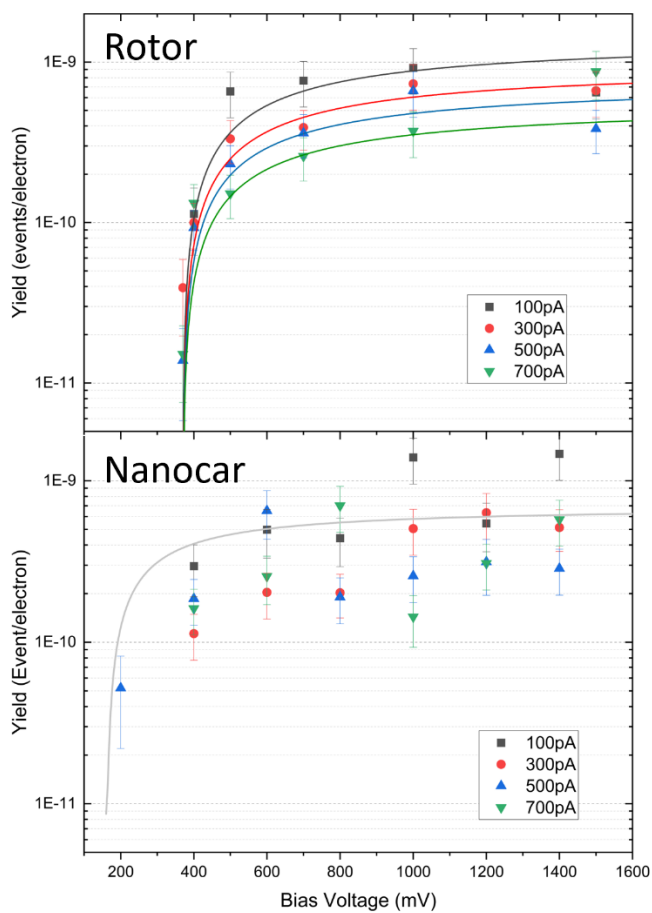

**Figure 6.** Yield (motion probability per electron) versus bias voltage for DMNI-P rotor (top) and nanocar (bottom) at pulses of fixed tunnelling current. The continuous lines for the rotor case (top) have been obtained fitting the experimental action spectra by the theory of Ref.[34] obtaining a threshold at 370 meV. In the lower panel, the continuous line is a simulation of the action spectra of the nanocar considering a threshold voltage of 165 meV and an overall yield constant of 3.5 x $10^{-10}$ events/electron. The fitting of the data at different currents was not possible because of the data spread.

At lower voltage, however, a translation becomes more favorable than a corresponding rotation. This indicates that low-energy vibrations are sufficient for inducing diffusion in the non-anchored physisorbed conformation, while higher energy is needed to overcome the rotational barrier of the anchored chemisorbed conformation, which can be reached by exciting the C-H stretch mode. Calculations of the vibrational modes of DMNI-P (see Figure S12), show similar modes for both rotor and nanocar and confirm the onset of the C-H stretch mode at 370 meV and a mixing of lower vibrational modes, including the C-C and C-N stretch between 155 meV and 167 meV. The continuous line in the lower panel of Figure 6, shows a simulation for a threshold voltage of 165 meV, in good agreement with the experimental data.

We therefore conclude that inelastic tunneling electrons are responsible for both rotation and translation, while for translation lower energy modes are already effective.

As already discussed in Ref.[11], the directionality of the movement (both as unidirectional rotation or directed translation) cannot be explained only by the inelastic excitation of vibrational modes, but is related to the virtual occupation of excited molecular electronic states during the electron tunneling through the molecule. In this specific case, the asymmetry of the molecule, its chirality on the surface, and the charge separation of the zwitterionic molecule in the electric field (which is always present between the metal tip and the metal surface during a voltage pulse) are encrypted in the asymmetry of the potential energy surfaces of the excited electronic states respect to the ground state, and are therefore contributing to the observed unidirectional motion.

CONCLUSIONS

We have presented a single-molecule machine showing two distinct functions depending on its adsorption conformation. Thanks to a coverage-related selectivity, we contemporarily observe rotors that are anchored by a fixed Au-O bond, and mobile nanocars that can controllably translate across the surface with atomic precision. The direction of rotation of the rotor is related to its chirality on the surface and can be inverted by manipulation, flipping the chirality without changing the anchoring position.

At bias voltages higher than 400 mV, the yields for rotation and translation are comparable. At lower voltage however, the nanocar is free to move laterally, while the rotation is blocked. This indicates that low-energy vibrational modes (*i.e.* C-C and C-N stretch modes at 150-170 meV) are sufficient to move the non-anchored physisorbed conformation, while higher energy modes (C-H stretch mode at 370 mV) are needed to induce the rotation of the chemisorbed anchored conformation.

METHODS

DMNI-P molecules were sublimated from a quartz crucible ($T_{evap}$ = 225 °C) on the clean Au(111) surface kept at room temperature under ultrahigh vacuum (UHV) conditions. The sublimation time was chosen between 2 – 7 sec depending on the target coverage. Before evaporation, the Au(111) single crystal was cleaned by subsequent cycles of $Ar^+$ sputtering and annealing to 723 K. STM experiments were performed using a custom-built instrument operating at a low temperature of T = 5 K under ultrahigh vacuum (p ≈ 1 × $10^{-10}$ mbar). All shown STM images were recorded in constant-current mode with the bias voltage applied to the sample.

Voltage pulse manipulations were performed by first positioning the STM tip above or near to the molecule and then by ramping up quickly the voltage bias. Both constant current and constant height modes were employed depending on the types of measurements. z(t) or I(t) curves, respectively, were recorded during the pulses, detecting the movement of the molecule. For constant height measurements, the tip-surface distance was calibrated by recording I(z) curves. STM images were recorded before and after the application of the pulse, determining the displacement of the molecule. Non-destructive tunneling parameters were selected (*e.g.* V = 0.2 V,

I = 5 pA) for acquiring STM images in order to lower the possibility to induce unexpected molecular movement during scanning.

All lateral manipulations were performed in constant-current mode. The lateral manipulation procedure involves three steps: (1) allowing the tip to vertically approach the molecules under a small bias and current to increase the tip–molecule interaction, (2) laterally driving the tip parallel to the surface in a precisely controlled trajectory, and (3) retracting the tip to normal scanning position. The STM captures images before and after each manipulation in all of the sequences.

The rotation or translation events were collected from the signal of tip height over time, where a sudden jump of signal can be observed. The corresponding yield (events/electron) calculations were done by the average of the population for all fixed voltage and current (n ≥ 10 for each point), and the statistical error was evaluated by standard deviation.

For geometry optimization and reaction path calculations, we used the DFT method as implemented in the CP2K software package (cp2k.org) with the Quickstep module[35]. We applied the Perdew-Burke-Ernzerhof exchange-correlation functional[36], the Goedecker-Teter-Hutter pseudo-potentials[37] and the valence double-ζ basis sets, in combination with the DFT-D2 method of Grimme[38] for van der Waals correction. We used 6 layers of gold, 3 upper layers allowed to be relaxed, planar supercell 29.8x19.9 Angstrom, vacuum size 40 Angstrom, maximum force 4.5x10-5 a.u. The data is analysed, and the images are made by the PyMOL Molecular Graphics System, Version 2.4 open-source build, Schrödinger, LLC.

The fitting of the action spectra of Figure 6 (rotor, upper panel) by the theory of Ref.[34] was done using the function:

$$Y = K \left[ \frac{2}{\pi}\left(1 - \frac{\hbar\Omega}{eV}\right)\left(\tan^{-1}\frac{2(eV - \hbar\Omega)}{\sigma} + \tan^{-1}\frac{2\hbar\Omega}{\sigma}\right) + \frac{\sigma}{2\pi eV}\log\frac{((\hbar\Omega)^2 + (\sigma/2)^2)}{((eV - \hbar\Omega)^2 + (\sigma/2)^2)} \right]$$

where *K* is the overall yield that controls the order of magnitude of the measured *Y*, the applied bias is *V*, $\Omega$ is the angular frequency of the vibrational mode mediating the reaction and $\sigma$ is the broadening of the reaction threshold. The fitting curves are then controlled by basically two parameters; $\Omega$ that fixes the threshold of a non-zero yield, and $\sigma$ that controls how fast the yield goes from zero to a measurable value.

The simulated curve of Figure 6 (nanocar, lower panel) was obtained fixing the threshold voltage (165 mV) and its spread (5 mV), and fitting the background yield with an overall constant of 3.5 x $10^{-10}$ events per electron.

ASSOCIATED CONTENT

**Supporting Information.** The following files are available free of charge:

- A supporting information PDF file is containing further experimental results and supplementary calculations.

- A mp4 movie (movie1.mp4) showing the changing of adsorption geometry during flipping, calculated by DFT.

- A mp4 movie (movie2.mp4) showing the movement of a rotor and a nanocar in the same image sequence.

AUTHOR INFORMATION

**Corresponding Author.** *Email: francesca.moresco@tu-dresden.de

**Author Contributions.** The manuscript was written through contributions of all authors. All authors have given approval to the final version of the manuscript.

ACKNOWLEDGMENTS

This work was funded by the European Union. Views and opinions expressed are however those of the authors only and do not necessarily reflect those of the European Union. Neither the European Union nor the granting authority can be held responsible for them.

This work has received funding from then European Innovation Council (EIC) under the project ESiM (grant agreement No 101046364) and the Horizon 2020 research and innovation program under the project MEMO, grant agreement no. 766864.

Support by the German Research Foundation (DFG) by the DFG Project 43234550, the Collaborative Research Centre (CRC) 1415, and the Initiative and Networking Fund of the German Helmholtz Association, Helmholtz International Research School for Nanoelectronic Networks NanoNet (VH-KO-606) is gratefully acknowledged. We thank the Center for Information Services and High Performance Computing (ZIH) at TU Dresden for computational resources.

F.L. thanks the Fonds der Chemischen Industrie (FCI) for a Liebig Fellowship and C.J. the WPI MANA project for financial support.

REFERENCES

1. Soe, W.-H.; Joachim, C., Towards a Molecular Mechanical Calculator. In *Single Molecule Mechanics on a Surface*, Moresco, F.; Joachim, C., Eds. Springer International Publishing: Cham, 2023, pp 141-156.
2. Wang, Z.; Hölzel, H.; Moth-Poulsen, K. Status and Challenges for Molecular Solar Thermal Energy Storage System Based Devices. *Chem. Soc. Rev.* **2022**, *51*, 7313-7326.
3. Ohmann, R.; Meyer, J.; Nickel, A.; Echeverria, J.; Grisolia, M.; Joachim, C.; Moresco, F.; Cuniberti, G. Supramolecular Rotor and Translator at Work: On-Surface Movement of Single Atoms. *ACS Nano* **2015**, *9*, 8394-8400.
4. Erbas-Cakmak, S.; Leigh, D. A.; McTernan, C. T.; Nussbaumer, A. L. Artificial Molecular Machines. *Chem. Rev.* **2015**, *115*, 10081-10206.

5. Aprahamian, I. The Future of Molecular Machines. *ACS Central Science* **2020,** *6*, 347-358.
6. Kay, E. R.; Leigh, D. A.; Zerbetto, F. Synthetic Molecular Motors and Mechanical Machines. *Angew. Chem. Int. Ed.* **2007,** *46*, 72-191.
7. Simpson, G. J.; Grill, L., Unidirectional Motion of Single Molecules at Surfaces. In *Single Molecule Mechanics on a Surface*, Moresco, F.; Joachim, C., Eds. Springer International Publishing: Cham, 2023, pp 1-27.
8. Stipe, B. C.; Rezaei, M. A.; Ho, W. Inducing and Viewing the Rotational Motion of a Single Molecule. *Science* **1998,** *279*, 1907-1909.
9. Perera, U. G. E.; Ample, F.; Kersell, H.; Zhang, Y.; Vives, G.; Echeverria, J.; Grisolia, M.; Rapenne, G.; Joachim, C.; Hla, S. W. Controlled Clockwise and Anticlockwise Rotational Switching of a Molecular Motor. *Nat. Nanotechnol.* **2013,** *8*, 46-51.
10. Tierney, H. L.; Murphy, C. J.; Jewell, A. D.; Baber, A. E.; Iski, E. V.; Khodaverdian, H. Y.; McGuire, A. F.; Klebanov, N.; Sykes, E. C. H. Experimental Demonstration of a Single-Molecule Electric Motor. *Nat. Nanotechnol.* **2011,** *6*, 625-629.
11. Eisenhut, F.; Kühne, T.; Monsalve, J.; Srivastava, S.; Ryndyk, D. A.; Cuniberti, G.; Aiboudi, O.; Lissel, F.; Zobač, V.; Robles, R.; Lorente, N.; Joachim, C.; Moresco, F. One-Way Rotation of a Chemically Anchored Single Molecule-Rotor. *Nanoscale* **2021,** *13*, 16077-16083.
12. Stolz, S.; Gröning, O.; Prinz, J.; Brune, H.; Widmer, R. Molecular Motor Crossing the Frontier of Classical to Quantum Tunneling Motion. *Proc. Natl. Acad. Sci.* **2020,** *117*, 14838-14842.
13. Pawlak, R.; Meier, T.; Renaud, N.; Kisiel, M.; Hinaut, A.; Glatzel, T.; Sordes, D.; Durand, C.; Soe, W.-H.; Baratoff, A.; Joachim, C.; Housecroft, C. E.; Constable, E. C.; Meyer, E. Design and Characterization of an Electrically Powered Single Molecule on Gold. *ACS Nano* **2017,** *11*, 9930-9940.
14. Nickel, A.; Ohmann, R.; Meyer, J.; Grisolia, M.; Joachim, C.; Moresco, F.; Cuniberti, G. Moving Nanostructures: Pulse-Induced Positioning of Supramolecular Assemblies. *ACS Nano* **2013,** *7*, 191-197.
15. Simpson, G. J.; García-López, V.; Petermeier, P.; Grill, L.; Tour, J. M. How to Build and Race a Fast Nanocar. *Nat. Nanotechnol.* **2017,** *12*, 604-606.
16. Manzano, C.; Soe, W. H.; Wong, H. S.; Ample, F.; Gourdon, A.; Chandrasekhar, N.; Joachim, C. Step-by-Step Rotation of a Molecule-Gear Mounted on an Atomic-Scale Axis. *Nat. Mater.* **2009,** *8*, 576-579.
17. Ren, J.; Freitag, M.; Schwermann, C.; Bakker, A.; Amirjalayer, S.; Rühling, A.; Gao, H.-Y.; Doltsinis, N. L.; Glorius, F.; Fuchs, H. A Unidirectional Surface-Anchored N-Heterocyclic Carbene Rotor. *Nano Letters* **2020,** *20*, 5922-5928.
18. Au Yeung, K. H.; Kühne, T.; Eisenhut, F.; Kleinwächter, M.; Gisbert, Y.; Robles, R.; Lorente, N.; Cuniberti, G.; Joachim, C.; Rapenne, G.; Kammerer, C.; Moresco, F. Transmitting Stepwise Rotation among Three Molecule-Gear on the Au(111) Surface. *J. Phys. Chem. Lett.* **2020,** *11*, 6892-6899.
19. Zhang, Y.; Calupitan, J. P.; Rojas, T.; Tumbleson, R.; Erbland, G.; Kammerer, C.; Ajayi, T. M.; Wang, S.; Curtiss, L. A.; Ngo, A. T.; Ulloa, S. E.; Rapenne, G.; Hla, S. W. A Chiral Molecular Propeller Designed for Unidirectional Rotations on a Surface. *Nat. Commun.* **2019,** *10*, 3742.

20. Jasper-Toennies, T.; Gruber, M.; Johannsen, S.; Frederiksen, T.; Garcia-Lekue, A.; Jäkel, T.; Roehricht, F.; Herges, R.; Berndt, R. Rotation of Ethoxy and Ethyl Moieties on a Molecular Platform on Au(111). *ACS Nano* **2020**, *14*, 3907-3916.
21. Shirai, Y.; Osgood, A. J.; Zhao, Y.; Kelly, K. F.; Tour, J. M. Directional Control in Thermally Driven Single-Molecule Nanocars. *Nano Lett.* **2005**, *5*, 2330-2334.
22. Kudernac, T.; Ruangsupapichat, N.; Parschau, M.; Maciá, B.; Katsonis, N.; Harutyunyan, S. R.; Ernst, K.-H.; Feringa, B. L. Electrically Driven Directional Motion of a Four-Wheeled Molecule on a Metal Surface. *Nature* **2011**, *479*, 208-211.
23. Kühne, T.; Au-Yeung, K. H.; Eisenhut, F.; Aiboudi, O.; Ryndyk, D. A.; Cuniberti, G.; Lissel, F.; Moresco, F. STM Induced Manipulation of Azulene-Based Molecules and Nanostructures: The Role of the Dipole Moment. *Nanoscale* **2020**, *12*, 24471-24476.
24. Grill, L.; Rieder, K. H.; Moresco, F.; Rapenne, G.; Stojkovic, S.; Bouju, X.; Joachim, C. Rolling a Single Molecular Wheel at the Atomic Scale. *Nat. Nanotechnol.* **2007**, *2*, 95-98.
25. Soe, W.-H.; Shirai, Y.; Durand, C.; Yonamine, Y.; Minami, K.; Bouju, X.; Kolmer, M.; Ariga, K.; Joachim, C.; Nakanishi, W. Conformation Manipulation and Motion of a Double Paddle Molecule on an Au(111) Surface. *ACS Nano* **2017**, *11*, 10357-10365.
26. Simpson, G. J.; García-López, V.; Daniel Boese, A.; Tour, J. M.; Grill, L. How to Control Single-Molecule Rotation. *Nat. Commun.* **2019**, *10*, 4631.
27. Joachim, C.; Rapenne, G. Molecule Concept Nanocars: Chassis, Wheels, and Motors? *ACS Nano* **2013**, *7*, 11-14.
28. 24 Hours of Toulouse. *Nat. Nanotechnol.* **2022**, *17*, 433-433.
29. Wei, P.; Menke, T.; Naab, B. D.; Leo, K.; Riede, M.; Bao, Z. 2-(2-Methoxyphenyl)-1,3-Dimethyl-1h-Benzoimidazol-3-Ium Iodide as a New Air-Stable N-Type Dopant for Vacuum-Processed Organic Semiconductor Thin Films. *J. Am. Chem. Soc.* **2012**, *134*, 3999-4002.
30. Lüssem, B.; Keum, C.-M.; Kasemann, D.; Naab, B.; Bao, Z.; Leo, K. Doped Organic Transistors. *Chem. Rev.* **2016**, *116*, 13714-13751.
31. Wei, P.; Oh, J. H.; Dong, G.; Bao, Z. Use of a 1h-Benzoimidazole Derivative as an N-Type Dopant and to Enable Air-Stable Solution-Processed N-Channel Organic Thin-Film Transistors. *J. Am. Chem. Soc.* **2010**, *132*, 8852-8853.
32. Schwarze, M.; Naab, B. D.; Tietze, M. L.; Scholz, R.; Pahner, P.; Bussolotti, F.; Kera, S.; Kasemann, D.; Bao, Z.; Leo, K. Analyzing the N-Doping Mechanism of an Air-Stable Small-Molecule Precursor. *ACS Appl. Mater. Interfaces* **2018**, *10*, 1340-1346.
33. Moresco, F. Manipulation of Large Molecules by Low-Temperature STM: Model Systems for Molecular Electronics. *Phys. Rep.* **2004**, *399*, 175-225.
34. Kim, Y.; Motobayashi, K.; Frederiksen, T.; Ueba, H.; Kawai, M. Action Spectroscopy for Single-Molecule Reactions – Experiments and Theory. *Prog. Surf. Sci.* **2015**, *90*, 85-143.
35. VandeVondele, J.; Krack, M.; Mohamed, F.; Parrinello, M.; Chassaing, T.; Hutter, J. Quickstep: Fast and Accurate Density Functional Calculations Using a Mixed Gaussian and Plane Waves Approach. *Comput. Phys. Commun.* **2005**, *167*, 103-128.
36. Perdew, J. P.; Burke, K.; Ernzerhof, M. Generalized Gradient Approximation Made Simple. *Phys. Rev. Lett.* **1996**, *77*, 3865-3868.
37. Goedecker, S.; Teter, M.; Hutter, J. Separable Dual-Space Gaussian Pseudopotentials. *Phys. Rev. B* **1996**, *54*, 1703-1710.
38. Grimme, S.; Antony, J.; Ehrlich, S.; Krieg, H. A Consistent and Accurate Ab Initio Parametrization of Density Functional Dispersion Correction (Dft-D) for the 94 Elements H-Pu. *J. Chem. Phys.* **2010**, *132*, 154104.

TABLE OF CONTENTS GRAPHIC

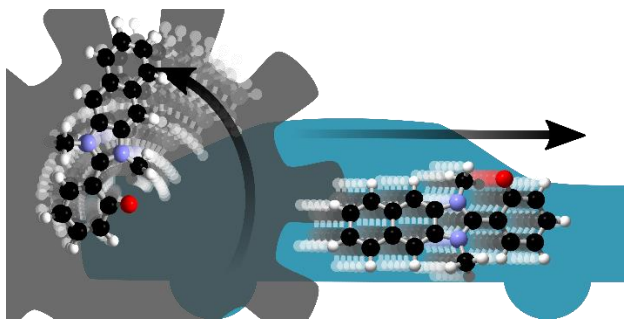